\documentstyle[prl,aps,epsfig,multicol]{revtex}
\begin{document}
\draft

\title{
\begin{flushright}
BNL-68068
\end{flushright}
Centrality Dependence of Charged Particle Multiplicity \\
        in  Au-Au Collisions at $\sqrt{s_{_{NN}}}=130$~GeV}


\author{
K.~Adcox,$^{40}$
S.{\,}S.~Adler,$^{3}$
N.{\,}N.~Ajitanand,$^{27}$
Y.~Akiba,$^{14}$
J.~Alexander,$^{27}$
L.~Aphecetche,$^{34}$
Y.~Arai,$^{14}$
S.{\,}
% ...
\\(PHENIX Collaboration)\\
}
\address{
% ...
$^{3}$Brookhaven National Laboratory, Upton, NY 11973-5000, USA\\
% ...
$^{14}$KEK, High Energy Accelerator Research Organization, Tsukuba-shi, Ibaraki-ken 305-0801, Japan\\
% ...
$^{27}$Chemistry Department, State University of New York - Stony Brook, Stony Brook, NY 11794, USA\\
% ...
$^{34}$SUBATECH (Ecole des Mines de Nantes, IN2P3/CNRS, Universite de Nantes) BP 20722 - 44307, Nantes-cedex 3, France\\
% ...
$^{40}$Vanderbilt University, Nashville, TN 37235, USA\\
% ...
}

\date{\today}        

\maketitle

\begin{abstract}
We present results for the charged-particle multiplicity distribution at
midrapidity in Au - Au collisions at $\sqrt{s_{_{NN}}}=130$~GeV measured
with the PHENIX detector at RHIC. For the 5\% most central collisions we
find $dN_{ch}/d\eta_{|\eta=0} = 622 \pm 1 (stat) \pm 41 (syst)$.  The
results, analyzed as a function of centrality, show a steady rise of the
particle density per participating nucleon with
centrality.  
\end{abstract}
\pacs{PACS numbers: 25.75.Dw}

\begin{multicols}{2}
\narrowtext

The Relativistic Heavy-Ion Collider (RHIC) at Brookhaven National
Laboratory started regular operation in June 2000, opening new frontiers
in the study of hadronic matter under unprecedented conditions of
temperature and energy density. The research is focused on the phase
transition associated with quark deconfinement and chiral symmetry
restoration expected to take place under those conditions.

In this letter, we report results for the charged-particle multiplicity
distribution at midrapidity in Au - Au collisions at
$\sqrt{s_{_{NN}}}=130$~GeV, as measured with the PHENIX detector. These
are the first RHIC results to span a broad impact parameter range.

Particle density at midrapidity is an essential global variable for the
characterization 
of high energy nuclear collisions,
providing information about the initial conditions, such as energy
density. The results presented here should help to constrain the wide
range of theoretical predictions \cite{bass-qm99} available at RHIC
energies and to discriminate among various mechanisms of entropy and
particle production.  In particular, we analyze the particle density as a
function of centrality, expressed by the number of participating nucleons.
Such an analysis may shed light on the relative importance of soft versus
hard processes of particle production and test the assumption of gluon
saturation expected at RHIC energies \cite{wang-gyulassy,ekrt}. 
Our results are compared to different models, to similar studies obtained
in Pb-Pb collisions at the CERN SPS \cite{wa98,wa97,na49-qm99} and to a
recent measurement performed at RHIC by PHOBOS \cite{phobos}.

The PHENIX detector is decribed in ref. 
\cite{phenix-detector}. The present analysis relies primarily on
three PHENIX subsystems: two layers of Pad Chambers (PC), called PC1 and PC3, 
used to determine the charged particle multiplicity,
the Zero Degree Calorimeters (ZDC) and the Beam-Beam Counters (BBC), used to derive 
the trigger and the off-line event selection.
The 
PC provide three-dimensional coordinates along the
charged-particle trajectories \cite{pc}. The two layers are mounted 
at radial distances of 2.49~m and 4.98~m, respectively, from the interaction
region.
Each 
layer has 8 wire chambers with cathode pad readout (see Fig. 1) and 
covers 90$^{o}$ in azimuth ($\phi$) and $\pm 0.35$ units of
pseudorapidity ($\eta$). 
The ZDC are small transverse-area hadron calorimeters that measure neutron
energy within a 2 mrad ($|\eta|>6$) cone around the beam direction
and are located at $\pm$18.25m from the center of the interaction region 
\cite{zdc}.
The BBC comprise two arrays of 64 photomultiplier tubes each equipped with
quartz Cherenkov radiators.  The BBC are located around the beam direction
at $\pm$1.44m from the center of the interaction region covering the full
2$\pi$ azimuth and the range $\eta = \pm$ (3.0 - 3.9) \cite{bbc}. 


The primary interaction trigger is generated by a coincidence between the
two BBC with at least two photomultipliers fired in each of
them and a requirement on the collision vertex position, usually $|z|\leq
20$~cm.  Based on detailed simulations of the BBC, this trigger reflects
[$92\pm2(syst)$]\% of the nuclear interaction cross section of 7.2 barns 
\cite{glauber}.  Another trigger is generated by a coincidence between the
two ZDC, each one with an energy signal larger than 10 GeV.
This trigger reflects the nuclear interaction plus the mutual Coulomb
dissociation cross sections.
Most BBC triggers (97.8\%)
also satisfy the ZDC trigger requirement.  The small percentage of
exclusive BBC triggers is due to 
inefficiencies of the ZDC trigger 
and background interactions.
These two sources are difficult to distinguish and we estimate a background event 
contamination of [$1\pm1(syst)$]\% of the total event sample.
 

This analysis is based on a sample of 137,784 events taken without magnetic field, 
satisfying the BBC trigger and with a reconstructed vertex position $|z|\leq 17$~cm. 


The number of primary charged particles per event is determined on a
statistical basis by correlating hits in PC1 and PC3.
The analysis procedure was developed and
corroborated by 
extensive simulations 
using the GEANT \cite{geant}
response of the PHENIX detector to events generated with HIJING 
\cite{hijing}.  
The vertex is reconstructed 
using the following algorithm: all hits in PC3 are combined with all hits
in PC1 and the resulting lines are projected onto a plane through the beam
line, 
perpendicular to the symmetry axis of the chambers (see Fig.~1). 
For 
events with more than $\sim$5 tracks, the distribution of these
projections along the Z axis produces a distinct peak which defines the
vertex position. For low-multiplicity events 
the vertex is reconstructed from the time difference between the 
two BBC.


Once the vertex is known,
all hits in PC3 are again combined
with all hits in PC1 and the resulting tracks are projected onto the plane
previously defined. The distribution of the distance R of the intersection
points to the vertex position, is shown in Fig.~2. 
This distribution contains real tracks and tracks from the
obvious combinatorial background inherent to the adopted procedure of
combining all hits in PC3 with all hits in PC1. The latter can be
determined by a mixed event technique; in the present analysis, each
sector in PC1 was exchanged with its neighbor and the resulting
combinatorial background is shown in Fig.~2 by the dotted line. The yield
of this background increases quadratically with R (leading to a linear
dependence in the differential dN/dR vs. R presentation of Fig.~2).


The R distribution of real tracks 
obtained by subtracting the
background (N$_B$) from the total number of tracks 
is shown in Fig.~2 by the dashed line. The sharp peak at small R is due to tracks from
primary particles originating at the vertex and the long tail is due to
decay products of primary particles decaying in flight.  In practice, the
track counting is performed up to a given R value. The fraction of counted
tracks as a function of R is obtained from the integral of the dashed
curve normalized to the total integral up to $R=\infty$. The tail 
at large R values is very well described by an exponential function
and therefore the extrapolation of the dashed curve to $R=\infty$ is
straightforward. The larger the value of R, the smaller is the correction
for the fraction of uncounted tracks but the larger the background to be
subtracted. Since the background can be reliably subtracted we performed
the track counting up to R=25cm, thus including 95.9\% of all tracks.


After subtracting the  background, a raw multiplicity
distribution is obtained to which we apply several corrections to
obtain the corrected distribution of primary charged
particles:


(i) A correction of 15.3\% accounts for inactive gaps between the
chambers, inactive electronic readout cards and dead pads in the PC1 and
PC3 detectors.

(ii)  Using cosmic rays, the pad chamber 
hit efficiency was measured to be 99.4\% for isolated single hits in agreement 
with an analytical study of the 
chamber performance \cite{analytical-calculation}.


(iii) The track losses due to the finite double hit resolution of the
chambers depend on the event multiplicity. The losses 
occur in the direct counting of tracks and in the
combinatorial background subtraction. The two effects were studied in
great detail with Monte Carlo techniques. For the first one, the
correction amounts to 13.3\% for the 5\% most central collisions, whereas
it is only 4.9\% for the 20-25\% bin. The second effect is a reduction of
the combinatorial background which leads to an increase in the track
multiplicity by $0.036\times N_B$. 


(iv) Finally there is a correction due to uncounted charged tracks. Two
sources contribute here. First, there is a correction of 4.3\% from tracks 
missed because the analysis is limited to R=25cm.  
The second source has two components. On the
one hand there are primary charged particles (mainly $\pi^\pm$) which
decay in flight. A large fraction of these decays is accounted for since,
as discussed above, they produce the tail of the R distribution in Fig.~2.
However, there is still a small number of decays which miss altogether PC1
and/or PC3 and those have to be added. On the other hand, there is
feed-down from neutral particle decays (mainly $K^o_S$ and $\pi^o$) which
lead to valid tracks. Those have to be subtracted. Both components depend
on particle composition and momentum distribution.  Lacking precise
information about them, we performed Monte Carlo simulations, applying to
the GEANT response of the pad chambers the same analysis procedure as to
real data and comparing the resulting multiplicity to the original
HIJING input.  The result is a net correction of only $\sim$2.8\%. This
correction is quite robust against changes in the input. For example, a
uniform 20\% increase in momentum $p$ of all pions reduces the net
correction to 1\%.
 
 
The systematic error in the multiplicity due to corrections (i) and (ii)
is estimated at less than 2\%. To these we have to add the 
errors
associated with the background subtraction, double hit resolution (which are multiplicity
dependent) and particle decays (which is multiplicity independent). We
estimate the first two uncertainties to be 4.6\% at the highest
multiplicities, based on Monte Carlo guidance and a comparison with an
identical analysis in which we imposed a 50\% larger double hit resolution
in 
PC1 and PC3.  For the
particle decay correction, we assign an
error of 4\% based on the effect of varying the momentum
distribution and the particle composition in the simulations and a
comparison with the results obtained with another data set measured while
the 
detectors were retracted by 44~cm from the nominal position.  Adding
these errors in quadrature results in a total systematic error of 6.5\% at
the highest multiplicities. An additional error in $dN_{ch}/d\eta$, which
enters in the analysis of the multiplicity versus centrality presented
below, is the uncertainty in the total number of events 
due to the
BBC trigger 
($\pm2\%$) and the 
event contamination
($\pm1\%$).


After applying the appropriate corrections we obtain in the lower panel of
Fig.~3 the minimum-bias charged-particle multiplicity distribution, in
the track acceptance 
$|\eta| <$ 0.35, $\delta\phi$ = 88.4$^o$. A factor of 5.82 thus converts
the observed number of tracks to $dN_{ch}/d\eta_{|\eta=0}$ in one unit of
pseudorapidity and full azimuth, yielding the lower horizontal scale in
Fig.~3.


Figure 3 also shows the multiplicity distributions for the four most
central bins, 0-5\% to 15-20\%. The bins were defined by cuts in the space
of BBC versus ZDC analog response (see Fig.~3 - upper panel) and refer to
percentiles of the total interaction cross section of 7.2 barns.  In order
to avoid the ambiguities inherent in the BBC vs. ZDC distribution, we
selected events with increasing centrality based upon the monotonic
response of 
another PHENIX detector.
The centroids of these events
projected onto the ZDC-BBC space determine the centrality contour
indicated by the solid line. The cuts are made perpendicular to this
contour. Simulations 
of the BBC and ZDC response were used to
account for the effect of physics and detector fluctuations in the
definition of these event classes and to relate them via a Glauber model
\cite{glauber} to the number of participating nucleons $N_{p}$ and 
of binary collisions $N_{c}$. The average charged particle density
scaled up to one unit of rapidity and the corresponding numbers $N_{p}$
and $N_{c}$ 
are tabulated in Table~1 for various centrality bins.  


The PHOBOS experiment has recently reported an average
$dN_{ch}/d\eta_{|\eta=0} = 555 \pm 12 (stat) \pm 35 (syst)$ for the 6\%
most central collisions \cite{phobos}. For the same centrality bin, we
find $dN_{ch}/d\eta_{|\eta=0} = 609 \pm 1 (stat) \pm 37 (syst)$. From the
preliminary results reported by NA49 in Pb-Pb collisions at
$\sqrt{s_{_{NN}}}=17.2$~GeV \cite{na49-qm99},
we derive a particle density $dN_{ch}/dy_{|y=0}= 410$.
Our result, $dN_{ch}/d\eta_{|\eta=0} = 622$ for the same centrality bin
(0-5\%), represents an increase of $\sim80\%$  after scaling it
by 1.2 to account for the transformation from $\eta$ to $y$.    


It has recently been emphasized \cite{wang-gyulassy} that the centrality
dependence of $dN_{ch}/d\eta$ allows one to discriminate between various
models of particle production.  We show in Fig.~4 our results
for $dN_{ch}/d\eta$ per 
participant pair as a function of the
number of 
participants $N_{p}$. Fig.~4 also shows the
{$p\overline p$} value at the same $\sqrt{s}$ taken from the UA5 analysis
\cite{ua5}. It is interesting to note that the extrapolation of our data
points to low multiplicities approaches the {$p\overline p$} value.
 

Models such as HIJING predict that there is a component of particle
production from soft interactions that scales linearly with $N_{p}$ and a
second component from hard processes (pQCD jets) that scales with 
N$_{c}$.  Following that, we fit the data of
Fig.~4 with the function $dN_{ch}/d\eta = A \times N_{p} + B \times N_{c}$
using the values of N$_{c}$ and N$_{p}$ tabulated in Table~1.  The 
results of the fit are shown in Fig.~4. Note that the errors in
A and B are anticorrelated. In 
such models, the values of A and B imply a large contribution of hard processes 
to particle production,
which increases with centrality from $\sim30\%$ at $N_{p}$=68 to
$\sim50\%$ for central collisions. HIJING predicts the same trend although
the calculated values are lower than the data by $\sim$ 15\%.
However, this is not a unique interpretation. 
At Alternating Gradient Synchrotron (AGS) energies, where hard
processes do not occur, the particle production per participant was also
observed to increase with centrality \cite{ahle}. At the CERN SPS energy
of $\sqrt{s_{_{NN}}}=17.2$~GeV, 
a similar behavior was observed. 
Using a different parametrization, $dN_{ch}/d\eta \propto
N_{p}^{\alpha}$, WA98 finds a best fit value of $\alpha = 1.07 \pm 0.04$ \cite{wa98}.
Experiment WA97 quotes a value $\alpha = 1.05 \pm 0.05$, consistent with 
the result of WA98 but also compatible with their assumption of
proportionality between multiplicity and participants \cite{wa97}.
A good fit to our data can also be obtained with this functional form
with a higher value of $\alpha = 1.16 \pm 0.04$.  One should note that
the CERN results are in the lab frame whereas our are in the center of
mass system.


Other models such as the EKRT \cite{ekrt} predict that at RHIC energies,
the large production of semi-hard gluons in a small volume may saturate
the gluon density. The resulting gluon fusion limits the total entropy
production and thus lowers the final particle production per participant.
The predictions of the EKRT model are also shown in Fig.~4.  We observe no
such saturation effect within our errors for Au + Au collisions at
$\sqrt{s_{_{NN}}}=130$~GeV, but instead see a steady rise in the particle
production per participant pair.

 
We thank the staff of the RHIC project, Collider-Accelerator, and Physics
Departments at BNL and the staff of PHENIX participating institutions for
their vital contributions.  We acknowledge support from the Department of
Energy and NSF (U.S.A.), Monbu-sho and STA (Japan), RAS, RMAE, and RMS
(Russia), BMBF and DAAD (Germany), FRN, NFR, and the Wallenberg Foundation
(Sweden), MIST and NSERC (Canada), CNPq and FAPESP (Brazil), IN2P3/CNRS
(France), DAE (India), KRF and KOSEF (Korea), and the US-Israel Binational
Science Foundation.

 

\vspace{-1cm}
\begin{figure}
\label{fig:geometry}
\centerline{\epsfig{file=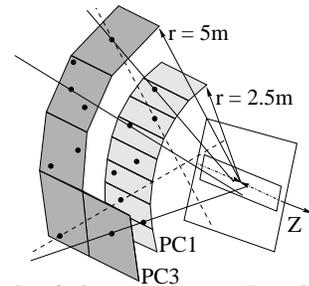,width=4cm}}
\caption {Sketch of the geometry.  For clarity, three 
PC3 sectors have been removed from the drawing.}
\end{figure}

\vspace{-0.5cm}
\begin{figure}
\label{fig:rpro}
\centerline{\epsfig{file=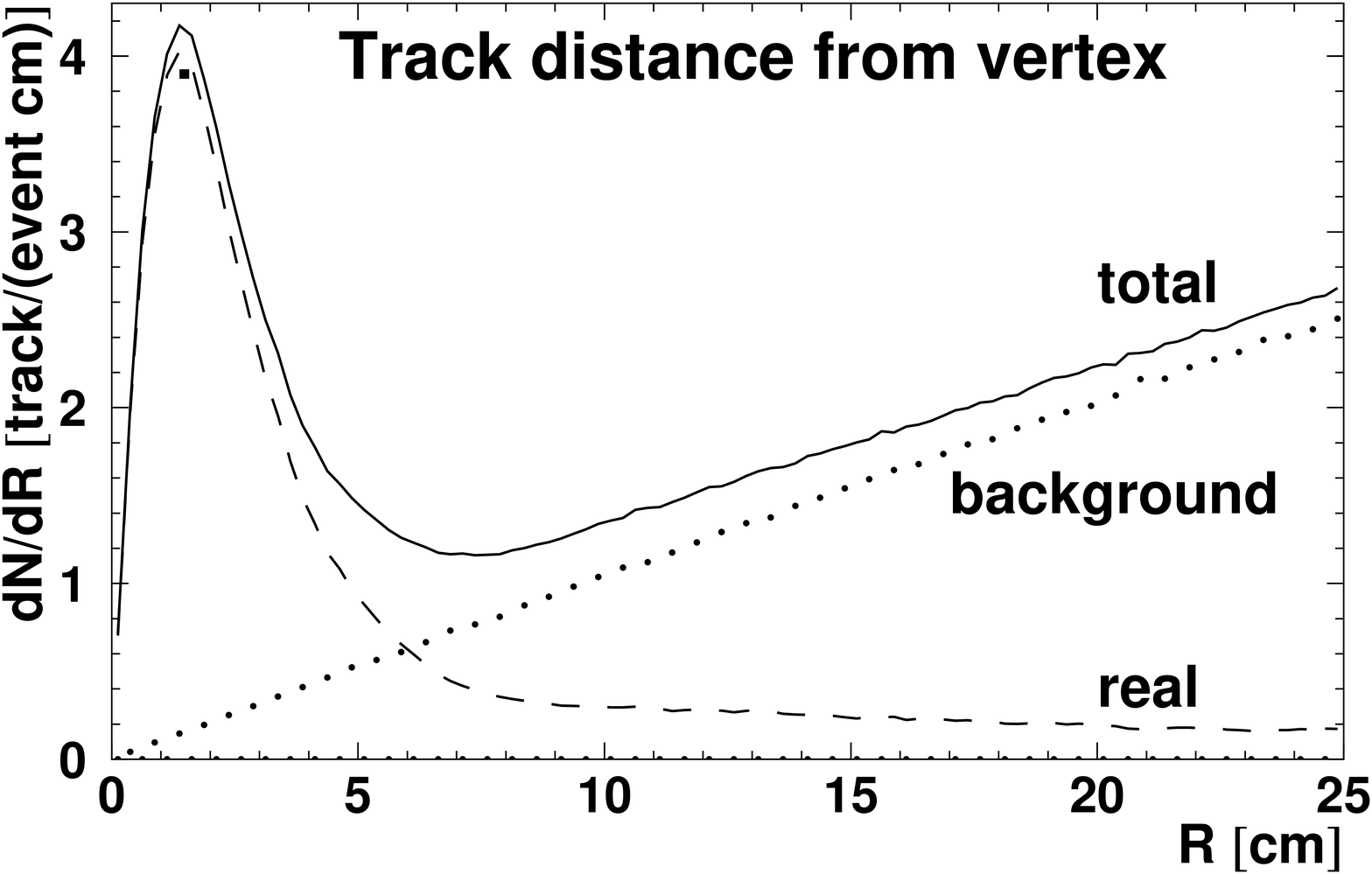,width=6cm}}
\caption {Tracks per event and per cm as a function of the distance to the event vertex.}
\end{figure}

\vspace{-0.5cm}
\begin{figure}
\label{fig:minbias}
\centerline{\epsfig{file=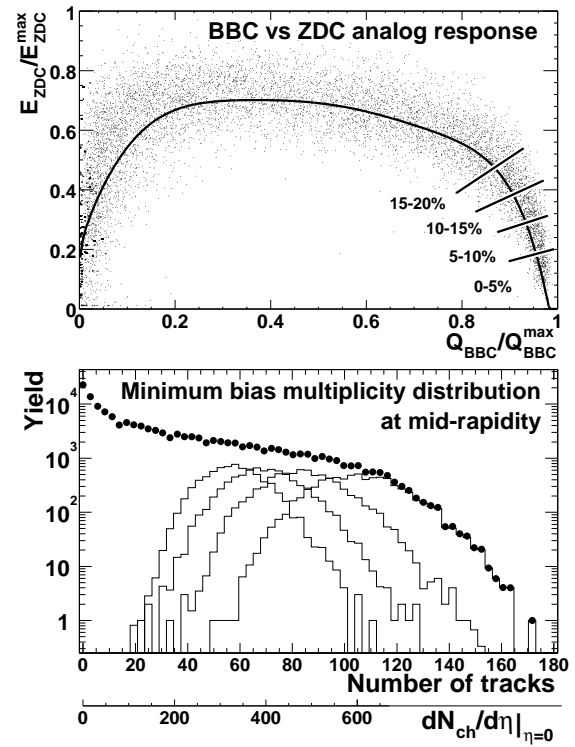,width=7.5cm}}
\caption {BBC vs. ZDC analog response (top panel) and minimum-bias
multiplicity distribution in the PHENIX measurement aperture (lower
panel). The lower axis converts the observed distribution to the
corresponding average $dN_{ch}/d\eta$ for track multiplicities less than
$\sim 120$; beyond that value the shape of the distribution has a
significant contribution from fluctuations into the measurement aperture.}
\end{figure} 
\vspace{-0.5cm}

\vspace{5cm}
\begin{figure}
\vspace{-0.3cm}
\centerline{\epsfig{file=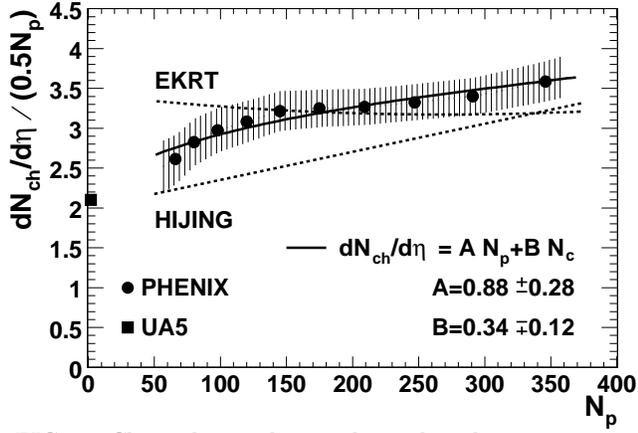,width=8.5cm}}
\caption {Charged-particle pseudorapidity density per 
participant pair  
vs. the number of 
participants.
Predictions from HIJING~\protect\cite{wang-gyulassy} and
EKRT~\protect\cite{ekrt} models, and a simple phenomenological fit are
also shown. The shaded area represents the systematic errors of
$dN_{ch}/d\eta$ and $N_{p}$. The errors of $N_{c}$ are in Table~1.}
\end{figure} 

\vspace{2cm}
\begin{table} 
\label{tab:participants}
\caption{Charged-particle density, number of 
participants and
collisions for various centrality bins expressed as percentiles of 
$\sigma_{geo}$=7.2 b.  The statistical
errors are negligible and only systematic errors are quoted.}
\begin{tabular}[]{cccc}
Centrality bin&$<dN_{ch}/d\eta_{|\eta=0}>$&$<N_{p}>$&$<N_{c}>$\\ \hline
 0 -  5 & 622 $\pm$ 41 & 347 $\pm$ 10 & 946 $\pm$ 146 \\
 5 - 10 & 498 $\pm$ 31 & 293 $\pm$ 9  & 749 $\pm$ 116 \\
10 - 15 & 413 $\pm$ 25 & 248 $\pm$ 8  & 596 $\pm$ 93  \\
15 - 20 & 344 $\pm$ 21 & 211 $\pm$ 7  & 478 $\pm$ 75  \\
20 - 25 & 287 $\pm$ 18 & 177 $\pm$ 7  & 377 $\pm$ 61  \\
25 - 30 & 235 $\pm$ 16 & 146 $\pm$ 6  & 290 $\pm$ 47  \\
30 - 35 & 188 $\pm$ 14 & 122 $\pm$ 5  & 226 $\pm$ 38  \\
35 - 40 & 147 $\pm$ 12 & 99  $\pm$ 5  & 170 $\pm$ 30  \\
40 - 45 & 115 $\pm$ 11 & 82  $\pm$ 5  & 130 $\pm$ 24  \\
45 - 50 & 89  $\pm$ 9  & 68  $\pm$ 4  & 101 $\pm$ 19  \\
\end{tabular}
\end{table}

\end{multicols}


\begin{references}
\bibitem{glauber} We used the Woods-Saxon nuclear density 
distribution, Au nucleus radius R=6.65$\pm$0.3fm, 
diffuseness a=0.54$\pm$0.01fm and
nucleon-nucleon cross section $\sigma_{nn}$=40$\pm$5mb.
\end{references}
\end{document}